\documentclass[aps,prd,twocolumn,floatfix,superscriptaddress,preprintnumbers]{revtex4}
\usepackage{amsfonts}
\usepackage{amssymb}
\usepackage{amsmath,mathtools}
\usepackage{ulem}
\usepackage{subfigure}
\usepackage{color}

\begin{document}

\title{Einstein Rings in Holography}
\author{Koji Hashimoto}
\affiliation{Department of Physics, Osaka University, Toyonaka, Osaka 560-0043, Japan}
\author{Shunichiro Kinoshita}
\affiliation{Department of Physics, Chuo University, Tokyo 112-8551, Japan}
\author{Keiju Murata}
\affiliation{Department of Physics, Osaka University, Toyonaka, Osaka 560-0043, Japan}
\affiliation{Department of Physics, College of Humanities and Sciences, Nihon University, Tokyo 156-8550, Japan}

\begin{abstract}
Clarifying conditions for the existence of a gravitational picture for a given quantum field theory (QFT) is 
one of the fundamental problems
in the AdS/CFT correspondence. 
We propose a direct way to demonstrate the existence of the dual black holes: {\it imaging an Einstein ring}.
We consider  
a response function of the thermal QFT on a two-dimensional sphere under a time-periodic localized source.
The dual gravity picture, if it exists, is a black hole in an asymptotic global AdS$_4$ and 
a bulk probe field with a localized source on the AdS boundary.
The response function corresponds to the asymptotic data of the bulk 
field propagating in the black hole spacetime.
We find a formula that converts the response function to the image 
of the dual black hole: 
The view of the sky of the AdS bulk from a point on the boundary.
Using the formula, we demonstrate that, for a thermal state dual to the Schwarzschild-AdS$_4$ spacetime,
the Einstein ring is constructed from the response function.
The evaluated Einstein radius is found to be determined by the total energy of the dual QFT.
Our theoretical proposal opens a door to gravitational phenomena on strongly correlated materials. 
\end{abstract}


\maketitle

\setcounter{footnote}{0}

\noindent

\section{Introduction}
One of the definitive goals of the research of the holographic principle,
or the AdS/CFT correspondence~\cite{Maldacena:1997re,Gubser:1998bc,Witten:1998qj},
is to find what class of quantum field theories (QFTs) or quantum materials possesses 
their gravity dual. 
Is there any direct test for the existence of a gravity dual for a given material?

\begin{figure}
\begin{center}
\includegraphics[scale=0.6]{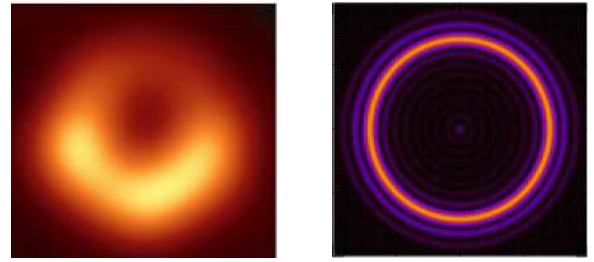}
\end{center}
 \caption{
(Left) Image of the black hole in M87 (This figure is taken from Ref.~\cite{EHT}.)
(Right) Image of the AdS black hole constructed from the response function.
 }
 \label{EHTfig}
\end{figure}

Among various gravitational physics, one of the most peculiar astrophysical 
objects is the black hole.
Gravitational lensing~\cite{Einstein} is one of the fundamental phenomena of strong
gravity.
When there is a light source behind a gravitational body, 
observers will see the Einstein ring.
If the gravitational body is a black hole, some light rays are so strongly
bent that they can go around the black hole many times, and especially
infinite times on the photon sphere.
As a result, multiple Einstein rings corresponding to winding numbers of
the light ray orbits emerge and infinitely concentrate
on the photon sphere.
Recently, the Event Horizon Telescope (EHT)~\cite{EHT}, which is an observational
project for imaging black holes, has captured the first image of the 
supermassive black hole in M87.
(See the left panel of Fig.~\ref{EHTfig}. 
We include the image of M87 for motivational purposes;
the physical origin of it, which may not be the same as that of
ours, is yet to be confirmed, though in Ref.~\cite{EHT}
it was claimed that it is consistent theoretically with a black hole
shadow surrounded by the brightest photon sphere.)
In this Letter, we propose a direct method to check the existence of a gravity dual from measurements in a given thermal QFT---{\it imaging the dual black hole as an Einstein ring}. 

\begin{figure}
\begin{center}
\includegraphics[scale=0.4]{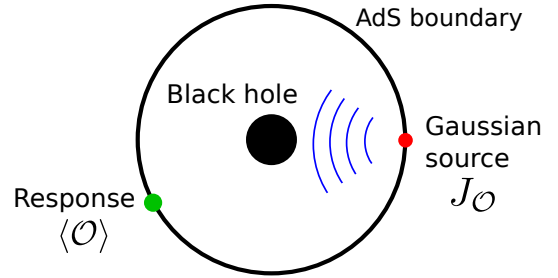}
\end{center}
\caption{
Our setup for imaging a dual black hole, the Schwarzschild-AdS$_4$ spacetime. 
An oscillating Gaussian source 
$J_\mathcal{O}$
is applied at a point on the AdS boundary.
Its response $\langle {\cal O}(x)\rangle$ is observed at another point on the boundary.
}
\label{setupfig}
\end{figure}

We demonstrate explicitly construction of holographic ``images'' of the dual black hole 
from the response function of the boundary QFT with external sources, as follows.
As the simplest example, we consider a $(2+1)$-dimensional 
boundary conformal field theory on a 2-sphere $S^2$ at a finite temperature,
and study 
a one-point function of a scalar operator ${\cal O}$,
under a time-dependent localized 
Gaussian 
source $J_{\cal O}$
with the frequency $\omega$.
We measure the local response function $e^{-i\omega t}\langle {\cal O}(\vec{x})\rangle$.
This QFT setup may allow a gravity dual (see Fig.~\ref{setupfig}), which is 
a black hole in the global AdS$_4$ and a probe
massless bulk scalar field in the spacetime.
The time-periodic source $J_{\cal O}$ 
amounts to a dynamical AdS boundary condition for the scalar field,
which injects a bulk scalar wave 
into the bulk from the AdS boundary.
The scalar wave propagates inside the black hole spacetime 
and reaches other points on the $S^2$ of the AdS boundary.
The scalar amplitude there corresponds to 
$\langle {\cal O}(\vec{x})\rangle$ in the QFT.

Using wave optics,
we find a formula which converts the response function $\langle {\cal O}(\vec{x})\rangle$ to the image 
of the dual black hole $|\Psi_\mathrm{S}(\vec{x}_\mathrm{S})|^2$ on a virtual screen:
\begin{equation}
 \Psi_\mathrm{S}(\vec{x}_\mathrm{S})= \int_{|\vec{x}|<d}d^2x\, \langle {\cal O}(\vec{x})\rangle e^{-\frac{i\omega}{f}\vec{x}\cdot\vec{x}_\mathrm{S}}\ ,
\label{lenstrans0}
\end{equation}
where $\vec{x}=(x,y)$ and $\vec{x}_\textrm{S}=(x_\textrm{S},y_\textrm{S})$ are Cartesian-like coordinates on boundary $S^2$ and 
the virtual screen, respectively, 
and we have set the origin of the coordinates to a given observation point.
This operation is mathematically 
a Fourier transformation of the
response function on a small patch with the radius $d$
around the observation point, 
that is, applying an appropriate window function.
Note that $f$ describes magnification of the image on the screen.
In wave optics, we have virtually used a lens with the focal
length $f$ and the radius $d$ to form the image.
The right panel of Fig.~\ref{EHTfig} shows a typical image of the AdS black hole computed
from the response function through our method.
The AdS/CFT calculation clearly gives a ring similar to the
observed image of the black hole by EHT. 
Equation~(\ref{lenstrans0}) can be regarded as the dual quantity of the Einstein ring.

Several criteria for QFTs to have a gravity dual have been proposed in some previous works.
A popular criterion is a large gap in the anomalous dimension spectra of CFT operators~\cite{Heemskerk:2009pn}.
The strong redshift of the black hole 
has been also used as the condition for the existence of gravity 
dual~\cite{Shenker:2013pqa,Maldacena:2015waa,Kitaev-talk}.
However, in the literature, it has never been discussed whether and how
we can holographically observe
effects of the gravitational lensing from observables in QFT.
We propose that the observation of the Einstein ring can be an alternative criterion 
for the existence of the gravity dual.
The method is simple and can be applied to any QFT on a sphere, thus probing efficiently a black hole
of its possible gravity dual. 
Once we have a strongly correlated material on $S^2$, we can apply 
a localized external source such as electromagnetic waves and measure its response in principle.
Then, from Eq.~(\ref{lenstrans0}), 
we would be able to construct the image of the dual black hole if it exists.
The holographic image of black holes in a material, if observed by a tabletop experiment, may serve as a
novel entrance to the world of quantum gravity.

\section{Scalar field in Schwarzschild-AdS$_4$ spacetime}
We consider Schwarzschild-AdS$_4$ (Sch-AdS$_4$) with the spherical horizon 
\begin{align}
ds^2=-F(r)dt^2+\frac{dr^2}{F(r)}+r^2(d\theta^2+\sin^2\theta d\varphi^2)\ ,
\label{SchAdS4}
\end{align}
where $F(r)=r^2/R^2+1-2 G M/r$, $R$ is the AdS curvature radius 
and $G$ is the Newton constant.
This is the black hole solution in the global AdS$_4$: The dual CFT is on $\mathbf{R}_t \times S^2$.
The event horizon is located at $r=r_\mathrm{h}$ defined by $F(r_\mathrm{h})=0$.
The mass $M$ is given by $M=r_\mathrm{h}(r_\mathrm{h}^2+R^2)/(GR^2)$,
which relates to total energy of the system in the dual CFT.
In what follows, we take the unit of $R=1$.

We focus on dynamics of a massless scalar field $\Phi(t,r,\theta,\varphi)$ in the Sch-AdS$_4$ satisfying 
the Klein-Gordon equation $\Box \Phi=0$. 
Near the AdS boundary ($r=\infty$), 
the asymptotic solution of the scalar field becomes
\begin{equation}
 \Phi=J_\mathcal{O}
-\frac{1}{2r^2}(\partial_t^2 - D^2)J_\mathcal{O}
+\frac{\langle \mathcal{O}\rangle }{r^3} +\cdots \ ,
\label{phiinf0}
\end{equation}
where $D^2$ is the scalar Laplacian on unit $S^2$.
In the asymptotic expansion, we have two independent functions, $J_\mathcal{O}$ and $\langle \mathcal{O}\rangle$,  
which depend on boundary coordinates $(t,\theta,\varphi)$.
In the AdS/CFT, the leading term $J_\mathcal{O}$ corresponds to the scalar source in the boundary CFT. 
On the other hand, the subleading term $\langle \mathcal{O}\rangle$ 
corresponds to its response function~\cite{Klebanov:1999tb}.

We consider that an axisymmetric and monochromatically oscillating Gaussian source is
localized at the south pole ($\theta=\pi$) of the boundary $S^2$: 
$J_\mathcal{O}(t,\theta,\varphi)=e^{-i\omega t} g(\theta)$, 
where
$g(\theta)=\exp[-(\pi-\theta)^2/(2\sigma^2)]/(2\pi \sigma^2)$ 
and $\sigma\ll \pi$.
In the bulk point of view, this source $J_\mathcal{O}$
is the boundary condition of the scalar field at the AdS boundary.
We also impose the ingoing boundary condition on the horizon of the Sch-AdS$_4$. 
Imposing these boundary conditions, we solve the Klein-Gordon equation numerically and 
determine the solution of the scalar field in the bulk.
We read off the response from the coefficient of $1/r^3$. 
Because of the symmetries of the source $J_\mathcal{O}$, the response function does not depend on
$\varphi$ and its time dependence is just given by $e^{-i\omega t}$.
Hence, we can write the response function as 
$\langle \mathcal{O}(t,\theta,\varphi)\rangle=e^{-i\omega t}\langle \mathcal{O}(\theta)\rangle$.

\begin{figure}
\begin{center}
\includegraphics[scale=0.7]{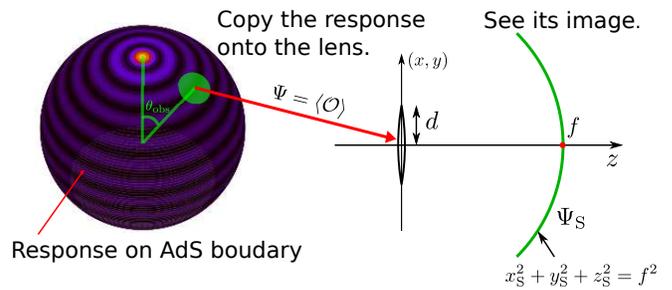}
\end{center}
\caption{
How to construct the image of the AdS black hole.
}
\label{restoim}
\end{figure}

\section{Imaging AdS black holes}
Figure~\ref{restoim} shows our procedure to obtain the image of the AdS black hole from the response function.
The sphere of the AdS boundary is depicted at the left side of the figure.
We show the absolute square of the response function $\langle \mathcal{O}(\theta)\rangle$ on the sphere
as the color map.
The brightest point is the north pole, i.e., the antipodal point of the Gaussian source.
We can observe the interference pattern resulting from the diffraction of the scalar wave by the black hole.
However, we cannot yet recognize this pattern directly as an 
``image of the black hole.''
To obtain the image of the black hole, we need to look at the response
function through a virtual ``optical system'' with a convex lens.

Let us
define an ``observation point'' at
$(\theta,\varphi)=(\theta_\textrm{obs},0)$ on the AdS boundary, where an
observer looks up into the AdS bulk.
To make an optical system virtually, 
we 
consider the flat three-dimensional space $(x,y,z)$ 
as shown in the right side of Fig.~\ref{restoim}.
We set the convex lens on the $(x,y)$-plane.
The focal length and radius of the lens will be denoted by $f$ and $d$.
We also prepare the spherical screen at $(x,y,z)=(x_\mathrm{S},y_\mathrm{S},z_\mathrm{S})$ 
with $x_\mathrm{S}^2+y_\mathrm{S}^2+z_\mathrm{S}^2=f^2$. 
We copy the response function around the observation point
onto the lens as the incident wave function 
and observe the image
formed on the screen.

We map the response function defined on $S^2$ onto the lens as follows.
We introduce new polar coordinates $(\theta',\varphi')$ as 
$\sin\theta'\cos\varphi'+i\cos\theta'=e^{i\theta_\textrm{obs}}(\sin\theta\cos\varphi+i\cos\theta)$,
such that the direction of the north pole is rotated to align with the
observation point: 
$\theta'=0 \Longleftrightarrow (\theta,\varphi)=(\theta_\textrm{obs},0)$.
Then, we define Cartesian coordinates as 
$(x,y)=(\theta'\cos\varphi',\theta'\sin \varphi')$
on the AdS boundary $S^2$.
In this coordinate system, we regard the response function around the observation point 
as the wave function on the lens: $\Psi(\vec{x})=\langle \mathcal{O}(\theta)\rangle$. 
For $\omega d \gg 1$, 
it is known that the image on the screen 
$\Psi_\mathrm{S}(\vec{x}_\mathrm{S})$ is obtained by 
the Fourier transformation of the incident wave $\Psi(\vec{x})$ within a
finite domain on the lens~\cite{Optics}.
(See also Refs.~\cite{Nambu:2012wa,Kanai:2013rga,Nambu:2015aea}.) 
This 
leads us to Eq.~(\ref{lenstrans0}) for imaging the AdS black hole.

\begin{figure*}
\begin{center}
\includegraphics[scale=0.9]{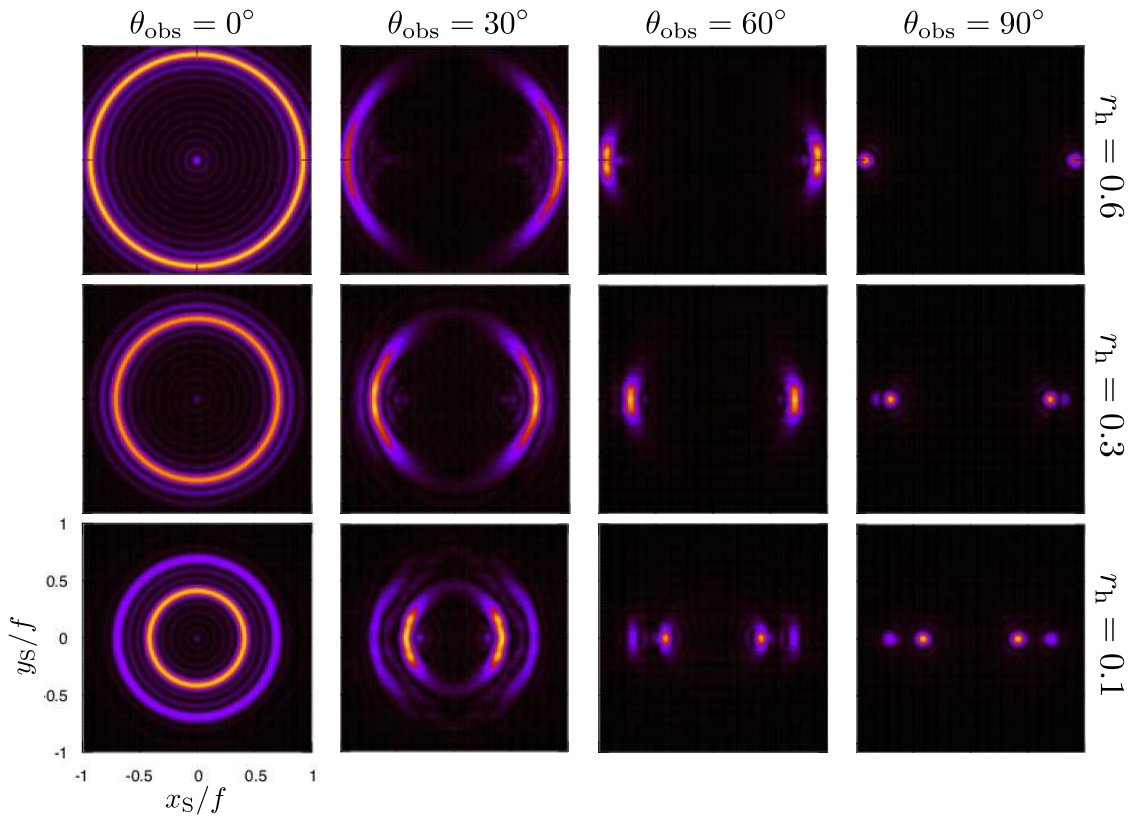}
\end{center}
 \caption{
Image of the Sch-AdS black hole for $\omega=80$, $\sigma=0.01$ and $d=0.5$.
The horizon radius and the observation point are varied as $r_\mathrm{h}=0.6,0.3,0.1$, and 
$\theta_\textrm{obs}=0^\circ,30^\circ,60^\circ,90^\circ$.
}
 \label{rh010306_view}
\end{figure*}

We now summarize our results. 
Figure~\ref{rh010306_view} shows images of the black hole observed at 
various observation points: 
$\theta_\textrm{obs}=0^\circ, 30^\circ, 60^\circ, 90^\circ$. 
The horizon radius is varied as $r_\mathrm{h}=0.6, 0.3, 0.1$.
We fix other parameters as $\omega=80$, $\sigma=0.01$ and $d=0.5$.
For $\theta_\textrm{obs}=0^\circ$, a clear ring is observed. 
As we will see 
later, 
this ring corresponds to
the light rays from the vicinity of the photon sphere of the Sch-AdS$_4$. 
Not only the brightest ring, we can also see some concentric striped patterns. 
They are caused by a diffraction effect with imaging, which is not directly
related to properties of the black hole, because we find these patterns change
depending on the lens radius $d$ and frequency $\omega$.
As we change the angle of the observer, the ring is deformed. 
We observed similar images as those for the asymptotically flat black hole~\cite{Nambu:2012wa,Kanai:2013rga,Nambu:2015aea}.
In particular, at $\theta_\textrm{obs}\sim 90^\circ$, we can observe a double image of the point source.
They correspond to light rays which are moving clockwise and anticlockwise,
winding around the black hole on the plane of $\varphi=0,\pi$.
The size of the ring becomes bigger as the horizon radius grows.

\section{Einstein radius}
\begin{figure}
\begin{center}
\includegraphics[scale=0.4]{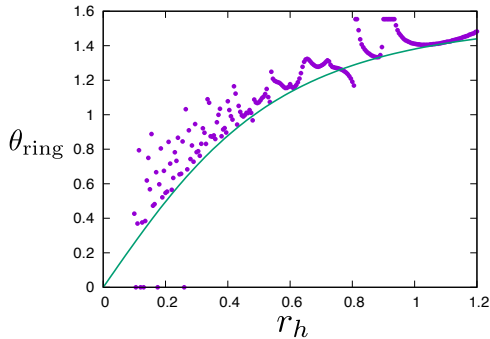}
\end{center}
\caption{
The Einstein radius $\theta_\textrm{ring}$ as a function of the horizon radius $r_\mathrm{h}$.
The green curve expresses the Einstein radius of the photon sphere, which is obtained by the geodesic approximation.
}
\label{thetaE}
 \end{figure}

To elucidate the property of the brightest ring, 
we focus on the observation point at $\theta_\textrm{obs}=0$ and search 
$x_\mathrm{S}=x_\textrm{ring}$ at which $|\Psi_\mathrm{S}(\vec{x}_\mathrm{S})|^2$ has maximum value.
(Since the image has the rotational symmetry in $(x_\mathrm{S},y_\mathrm{S})$-plane for $\theta_\textrm{obs}=0$, 
we only consider the $x_\mathrm{S}$-axis.)
We will refer to
the angle of the Einstein ring  $\theta_\textrm{ring}=\sin^{-1}(x_\textrm{ring}/f)$ as the Einstein radius.
Figure~\ref{thetaE} shows the Einstein radius $\theta_\textrm{ring}$ by the purple points 
varying the horizon radius $r_\mathrm{h}$.
Although the Einstein radius fluctuates as the function of $r_\mathrm{h}$, 
it has an increasing trend as $r_\mathrm{h}$ is enlarged.

From geometrical optics, 
there is an infinite number of
null geodesics connecting antipodal points on the AdS boundary, which are
labeled by the winding number $N_\textrm{w}$:
the number of times a geodesic goes around the black hole.
Each geodesic will form an Einstein ring whose radius is determined by
the angle of incidence to the AdS boundary.
Which geodesic in the geometrical optics corresponds to the ring found
in the image in the wave optics?
In the Sch-AdS$_4$ spacetime~(\ref{SchAdS4}), there is the photon sphere ($N_\textrm{w}=\infty$) at 
$r= 3r_\mathrm{h}(r_\mathrm{h}^2+1)/2$.
Solving the null geodesic equation from the photon sphere, we obtain the angle of incidence  
of the geodesic to the AdS boundary $\vartheta_\mathrm{i}$ as
\begin{equation}
 \sin \vartheta_\mathrm{i} = \frac{r_\mathrm{h}(r_\mathrm{h}^2+1)}{(r_\mathrm{h}^2+1/3)\sqrt{r_\mathrm{h}^2+4/3}}\ .
\end{equation}
This geodesic prediction for the Einstein radius of the photon sphere
is shown by the green curve in Fig.~\ref{thetaE}.
The curve seems consistent with the Einstein radius from the wave optics.
This indicates that the major contribution to the brightest ring in the
image is originated by the ``light rays`` from the vicinity of the photon sphere, which
are infinitely accumulated.
Although there 
are expected to
be multiple Einstein rings corresponding to light lays with winding
numbers $N_\textrm{w}=0,1,2,\ldots$, the contribution for the image from small $N_\textrm{w}$ 
may be
so small that 
we cannot 
resolve 
them within our numerical accuracy.

The deviation of the Einstein radius $\theta_\textrm{ring}(r_\textrm{h})$ 
from the geodesic prediction can be considered as some wave effects.
In the AdS cases, whether the geometrical optics can adapt to imaging of
black holes is not so trivial even for a large value of $\omega$.
The Eikonal approximation, which supports the geometrical optics, will
inevitably break down 
since a component of the metric rapidly varies near the AdS boundary as $\sim r^2$. 
Our results based on the wave optics imply that the geometrical
optics is qualitatively valid but gives a non-negligible deviation even
for a large $\omega$.

When $\omega d \gg 1$, Eq.~(\ref{lenstrans0}) can be rewritten as
$
 \Psi_\textrm{S} \propto G_\ell(\omega)|_{\ell \simeq \omega\sin\theta_\textrm{S}}
$
by using the retarded Green's function $G_\ell(\omega)$ for frequency
$\omega$ and azimuthal quantum number $\ell$ of the spherical harmonics
on $S^2$.
(See appendix~\ref{EringGreen} for more details.)
It turns out that the poles of $G_\ell(\omega)$, which correspond to quasinormal modes
(QNMs) for the bulk black hole, play an important role for the image formation.
In particular, the Einstein radius $\theta_\textrm{ring}$ for a given
$\omega$ implies that
there should exist the quasinormal modes such that the real part of the
frequency is close to $\omega$ with respect to 
$\ell = \omega \sin\theta_\textrm{ring}$.
Since the QNMs are closely related to the unstable photon orbit in the
Eikonal limit $\omega \simeq \ell \gg 1$, this is consistent with our results.

Furthermore,
in appendix~\ref{EringGreen}, 
we estimate the ``Einstein radius'' for 
a weakly coupled 
$\phi^4$-theory with mass $m$, coupling $\lambda$ and temperature  $T$. 
At the one loop level, we have
\begin{equation}
 \theta_\textrm{ring}\simeq \frac{\pi}{2}-\frac{m_T}{\omega}\ ,
\end{equation}
where $m_T^2=m^2+\mathcal{O}(\lambda T)$ is the effective mass with the thermal correction.
The temperature dependence of $\theta_\textrm{ring}$ is suppressed for
$\omega\to \infty$, 
as opposed to the strongly coupled cases associated with gravitate dual.
This result suggests that, from the temperature dependence of the Einstein radius,  
we can diagnose if a given QFT has its gravity dual or not.

\section{Discussion}
We have shown 
that, if the dual black hole exists, 
we can construct 
the image of the AdS black hole from the observable in the thermal QFT.
In other words, being able to observe the image the AdS black hole in the thermal QFT
can be regarded as a necessary condition for the existence of the dual black hole.
Finding conditions for the existence of the dual gravity picture for a given quantum field theory is 
one of the most important problems 
in the AdS/CFT. 
We would be able to use the imaging of the AdS black hole as a test 
for it.
One of the possible  applications 
is  superconductors. 
It is known that properties of high $T_c$ superconductors can be captured by the black hole physics
in AdS~\cite{Gubser:2008px,Hartnoll:2008vx,Hartnoll:2008kx}.
One of the 
other interesting applications 
is the Bose-Hubbard model,
which at the quantum critical regime has been conjectured to have 
a gravity dual~\cite{Sachdev:2011wg,Fujita:2014mqa,BHMOTOC}. 
Experimental realization of such strongly correlated materials on a two-dimensional sphere 
is a possible entrance to the world of quantum gravity: 
Measurement of responses under localized sources on
such materials will lead to a novel observation of 
Einstein rings by tabletop experiments.

\begin{acknowledgments}
We would like to thank 
Vitor Cardoso, 
Paul Chesler, 
Yoshimasa Hidaka,
Sousuke Noda 
and
Chulmoon Yoo,
for useful discussions and comments.
We would also like to thank organizers and participants of the YITP workshop YITP-T-18-05 ``Dynamics in Strong Gravity Universe''
for the opportunity to present this work and useful comments.
The work of K.~H.~was supported 
in part by JSPS KAKENHI Grants No.~JP15H03658, 
No.~JP15K13483, and No.~JP17H06462. 
The work of K. M. was supported by JSPS KAKENHI Grant No. 15K17658 and 
in part by JSPS KAKENHI Grant No. JP17H06462.
The work of S.~K.~was supported in part by JSPS KAKENHI Grants No.~JP16K17704.
\end{acknowledgments}

\appendix 
\section{Einstein ring from retarded Green function}
\label{EringGreen}

The linear response $\langle \mathcal{O}\rangle$ with respect to the external source $J_\mathcal{O}$ on unit $S^2$
is written as 
\begin{multline}
 \langle \mathcal{O}(t,\theta)\rangle
= -2\pi\int dt' d\theta' \sin\theta' \\
\times 
G(t,t',\theta,\theta')J_\mathcal{O}(t',\theta')\ ,
\label{eq:response_func}
\end{multline}
where we have assumed that the source $J_\mathcal{O}$ is axisymmetric. 
We introduce the retarded Green function $G(t,t',\theta,\theta')$.
It is well known that the Green function is given by the real time correlation function as
\begin{equation}
 G(t,t',\theta,\theta')=-i \Theta(t-t') \langle [\mathcal{O}(t,\theta),\mathcal{O}(t',\theta')]\rangle\ ,
\end{equation}
where $\Theta(t-t')$ is the step function and $\langle \cdots\rangle$ is the ensemble average with the equilibrium density matrix.
(For example, see Ref.~\cite{Natsuume:2014sfa} for the derivation.)
Let us suppose that the source $J_\mathcal{O}$ is monochromatic with
a frequency $\omega$.
The Green function and the source can be expanded in terms of 
Fourier modes 
and spherical harmonics 
$Y_\ell(\theta)\equiv Y_{\ell\,m=0}(\theta)$ as 
\begin{align}
 &G(t,t',\theta,\theta')=\sum_\ell \int \frac{d\omega'}{2\pi} e^{-i\omega' (t-t')} \nonumber \\
&\hspace{3cm}\times G_\ell(\omega) Y_\ell(\theta)Y_\ell(\theta')\ ,\\
&J_\mathcal{O}(t',\theta')=e^{-i\omega t'}\sum_\ell J_\ell Y_\ell(\theta')\ .
\end{align}
Thus, we can rewrite the response function (\ref{eq:response_func}) as
\begin{equation}
 \langle \mathcal{O}(t,\theta)\rangle 
=-e^{-i\omega t} \sum_\ell  G_\ell(\omega) J_\ell Y_\ell(\theta)\ .
\label{OofG}
\end{equation}

In the main text of the letter, we proposed 
the formula for constructing the holographic image of the Einstein ring from the response function.
For simplicity, we set the observation point at the north pole:
$\theta_\textrm{obs}=0$, which is the antipodal point of the  
external source localized at $\theta = \pi$.  
Then, the image of the Einstein ring on the virtual screen is given by
\begin{multline}
 \Psi_S(t,\theta_\textrm{S})=\int_0^{2\pi} d\varphi \int_0^d d\theta\sin\theta\\
 \times \langle \mathcal{O}(t,\theta)\rangle 
\exp\left(-\frac{i\omega}{f}\vec{x}\cdot \vec{x}_S\right)\ ,
\label{Psi_S}
\end{multline} 
where $d$ is the radius of the lens and we assume $d\ll 1$ in the unit of the radius of $S^2$.
We have introduced polar coordinates on the boundary $S^2$ and the virtual screen as
\begin{equation}
\begin{split}
&\vec{x}=\sin\theta ( \cos\varphi, \sin\varphi)\ ,\\
&\vec{x}_S=f\sin\theta_S (\cos\varphi_S,\sin\varphi_S)\ .
\end{split}
\end{equation}
Note that the formula (\ref{Psi_S}) means the Fourier transform of
the response function multiplied by a window function which is nonzero
only within a small finite region on $S^2$.

Now, we will rephrase the formula in terms of the retarded Green
function.
We perform the integration with respect to $\varphi$ by using 
$\vec{x}\cdot \vec{x}_S = f\sin \theta \sin\theta_S \cos(\varphi-\varphi_S)$,
and plug Eq.~(\ref{OofG}) into Eq.~(\ref{Psi_S}).
As a result, we obtain
\begin{align}
 &\Psi_S(t,\vec{x}_S)= 
2\pi\int_0^d  d\theta \sin \theta \, \langle \mathcal{O}(t,\theta)\rangle \mathcal{J}_0(\omega \sin \theta_\textrm{S} \sin\theta)\nonumber\\
&= -2\pi e^{-i\omega t}  \sum_\ell  G_\ell(\omega) J_\ell \int_0^d  d\theta \sin \theta \nonumber\\
&\hspace{3cm}
\times Y_\ell(\theta) \mathcal{J}_0(\omega \sin \theta_\textrm{S} \sin\theta)\ ,
\label{PsiS2}
\end{align}
where $\mathcal{J}_n(x)$ is the Bessel function of the first kind, which comes from the $\varphi$-integration.
For $\theta \le d \ll 1$, the spherical harmonics can be approximated by the Bessel function as
\begin{equation}
 Y_\ell(\theta)\simeq \sqrt{\frac{\ell+1/2}{2\pi}}\mathcal{J}_0((\ell+1/2) \theta)\ .
\end{equation}
Using the above expression and replacing $\sin\theta\simeq \theta$ in Eq.~(\ref{PsiS2}), 
we can explicitly perform the $\theta$-integration and obtain
\begin{multline}
 \Psi_S(t,\vec{x}_S)
\simeq -\sqrt{2\pi}d^2 
e^{-i\omega t}  \sum_\ell  (\ell+1/2)^{1/2} \\
\times G_\ell(\omega) \, J_\ell \, 
\Delta\left((\ell+1/2)d,\omega\sin\theta_\textrm{S} d\right)\ ,
\label{PsiS3}
\end{multline}
where we have defined 
\begin{equation}
 \Delta(x,y)\equiv \frac{x\mathcal{J}_1(x)\mathcal{J}_0(y)-y\mathcal{J}_1(y)\mathcal{J}_0(x)}{x^2-y^2}\ .
\end{equation}
The function $\Delta(x,y)$ has the highest peak with width $\sim \pi$ at
$x=y$ and is damping as $|x-y|$ becomes large.
If $\omega d \gg 1$, 
$\Delta\left((\ell+1/2)d,\omega\sin\theta_\textrm{S} d\right)$ in 
Eq.~(\ref{PsiS3}) has
greater values around 
$|\ell/\omega - \sin\theta_\textrm{S}| \lesssim \pi/\omega d \ll 1$.
Thus, we can eventually evaluate 
\begin{equation}
 \Psi_S(t,\vec{x}_S)
\propto
e^{-i\omega t} G_\ell(\omega) J_\ell\bigg|_{\ell=\omega\sin\theta_\textrm{S}}  \ .
\end{equation}
Formally, Eq.~(\ref{PsiS3}) means that the image on the virtual screen
corresponds to convolution of the Green function and the window function
in wave-number space $\ell$.
Therefore, it turns out that the image resolution is characterized by
the window function as 
$\Delta \ell/\omega \simeq \Delta \theta_\textrm{S} \simeq \pi/\omega d$.

In the view of the gravity side, poles of the retarded Green function in
the frequency domain correspond to quasi-normal mode (QNM) frequencies, $\omega=\Omega_\ell^n\in \bf{C}$, where $n=0,1,2,\cdots$ represent overtone numbers.
We can expect that $|\Psi_S|^2$ has a large value when the frequency
of the given monochromatic source, $\omega$, is close to the position of a
QNM frequency in the complex $\omega$-plane, so that the Einstein ring
is formed on the screen.
In other words, the condition
for the Einstein radius $\theta_\textrm{S}=\theta_\textrm{ring}$ is written as
\begin{equation}
 \omega\simeq \textrm{Re}\, \Omega^n_{\ell=\omega \sin\theta_\textrm{ring}} \ ,
  \label{thetaring_cond}
\end{equation}
for an overtone number $n$.

The field equation of the massless scalar field, 
$\Phi(t,r,\theta,\varphi) \equiv e^{-i\omega t}\sum_\ell Y_\ell(\theta) \psi_\ell(r)/r$,
is given by 
\begin{multline}
 \bigg[
 - \frac{d^2}{dr_*^2} + \ell(\ell+1)v(r) \\
+ \frac{F(r)F'(r)}{r}
\bigg]\psi_\ell(r)
= \omega^2\psi_\ell(r) , 
\end{multline}
where $v(r) \equiv F(r)/r^2$
and 
we have introduced the tortoise coordinate $dr_* = dr/F(r)$.
According to the WKB analysis~\cite{Ferrari:1984zz},
QNMs are characterized by the behavior of the effective potential around
an extremum.
In the Eikonal limit $\omega \simeq \ell \gg 1$, the local maximum of
a part of the potential $v(r)$ plays a significant role and is given by 
\begin{equation}
\begin{split}
&v_\mathrm{max} \equiv v(r_\mathrm{max}) =
  \frac{(3r_\mathrm{h}^2 + 4)(3r_\mathrm{h}^2 + 1)^2}
  {27r_\mathrm{h}^2(r_\mathrm{h}^2 + 1)^2} ,\\
&r_\mathrm{max} \equiv \frac{3}{2}r_\mathrm{h}(r_\mathrm{h}^2 + 1) .
\end{split}
\end{equation}
As a result, the QNMs that originate from this local maximum are
described by 
\begin{equation}
\begin{split}
&\textrm{Re}\, \Omega_\ell^n = \sqrt{\ell(\ell+1)v_\textrm{max}-\alpha^2}\ ,\\
&\textrm{Im}\, \Omega_\ell^n = \alpha\left(k(n)+\frac{1}{2}\right)\ ,
\end{split}
\end{equation}
where $\alpha\equiv \sqrt{-(d^2v/dr_\ast^2)/(2v)}|_{r=r_\textrm{max}}$
and $k(n)$ is a real number of $O(1)$.
Even though this expression of the QNM frequencies is derived for asymptotically flat spacetime, this is still valid for asymptotically AdS case 
since the QNM is highly oscillating as a function of $r_\ast$ for
$\omega ,\ell\gg 1$ and can be easily connect to the desired solution near the AdS boundary.
(For detailed WKB analysis in asymptotically AdS spacetimes, see Refs.~\cite{Festuccia:2008zx,Berti:2009wx,Berti:2009kk,Dias:2012tq}.)
For $\ell \gg 1$, we have $\textrm{Re}\, \Omega_\ell^n \simeq \ell\sqrt{v_\textrm{max}}$ and Eq.~(\ref{thetaring_cond}) gives
\begin{equation}
 \sin\theta_\textrm{ring}\simeq \frac{1}{\sqrt{v_\textrm{max}}}\ .
\end{equation}
This is consistent with our direct numerical calculations.

The retarded Green function $G_\ell(\omega)$ is a well-studied quantity in quantum field theories.
For example,  the Green function of a weakly coupled field theory with mass $m$ and coupling $\lambda$
is given by
\begin{equation}
 G_\ell(\omega)=\frac{1}{-\omega^2+\ell(\ell+1)+m_T^2}\ ,
\end{equation}
where $m_T=m^2+\mathcal{O}(\lambda T^2)$ is the effective mass with the thermal effect. 
Then, the Einstein radius for weakly coupled theory is given by
$\sin^2\theta_\textrm{ring} = 1-m_T^2/\omega^2$. It does not depend on the temperature for a sufficiently large $\omega$ and gives $\theta_\textrm{ring}\simeq \pi/2$.
This suggests that, from the temperature dependence of the Einstein ring,
we can diagnose if a given quantum field theory has its gravity dual.

\end{document}